\documentclass{easychair}
\usepackage{xspace}
\usepackage{xcolor}
\usepackage{tikz}
\usepackage{textcomp}
\usetikzlibrary{shapes.geometric}
\definecolor{dblue}{rgb}{0.2,0.2,0.7}%
\hypersetup{%
  colorlinks=true,%
  pdfborder={0 0 0},%
  linkcolor=dblue,%
  citecolor=dblue,%
  urlcolor=dblue,%
}

\definecolor{mg}{HTML}{9E2BE6}
\definecolor{free}{HTML}{0E0EF4}
\definecolor{bound}{HTML}{0E810E}
\definecolor{mygreen}{HTML}{009966}
\definecolor{myblue}{HTML}{006699}
\definecolor{redish}{HTML}{FFC1C1}
\definecolor{yellowish}{HTML}{EEE8AA}

\newcommand\isahome{\texttt{ISABELLE\char`_HOME}}
\newcommand\isayear{2018}
\newcommand\isaversion{\texttt{Isabelle\isayear}\xspace}
\newcommand\pg{Proof General\xspace}
\newcommand\isa{Isa\-belle\xspace}
\newcommand\jedit{jEdit\xspace}
\newcommand\isajedit{\isa/\hspace{0pt}\jedit}
\newcommand\secref[1]{Section~\ref{sec:#1}}
\newcommand\figref[1]{Figure~\ref{fig:#1}}
\newcommand*\type[1]{{\tt#1}}
\newcommand\error[1]{%
 \begin{tikzpicture}[baseline=(X.base)]
 \node[fill=redish](X) at (0,0) {\texttt{#1}};
 \end{tikzpicture}
}
\newcommand\warning[1]{%
 \begin{tikzpicture}[baseline=(X.base)]
 \node[fill=yellowish](X) at (0,0) {\texttt{#1}};
 \end{tikzpicture}
}
\newcommand\inot{\raisebox{-4pt}{{\large\texttt{\char`~}}}}
\newcommand*\gr[1]{\textcolor{mygreen}{#1}}
\newcommand*\bl[1]{\textcolor{myblue}{#1}}
\newcommand*\mg[1]{\textcolor{mg}{#1}}
\newcommand*\free[1]{\textcolor{free}{#1}}
\newcommand*\bound[1]{\textcolor{bound}{#1}}
\newcommand\scut[1]{\textsf{#1}}
\newcommand\menu[1]{\textsf{#1}}
\newcommand\then{\textsf{$\to$}}

\newcommand\tvar[1]{\mg{\textquotesingle#1}}

\def\pghead{\lower2.75pt\vbox to 0pt{\vss\hbox{\includegraphics[width=12pt]{pghead}}\vss}}
  {\par\endgroup\medbreak}
\def\warnsym{\raise3pt\vbox to 0pt{\vss\hbox{%
\tikz \node [draw,thick,isosceles triangle,rounded corners=2pt,shape border
rotate=90,isosceles triangle stretches,inner sep=0pt,minimum height=3mm,minimum
width=12pt] {\raisebox{1pt}{\small\textbf{!}}};}\vss}}
\newenvironment{warn}{\medskip\medbreak\begingroup \clubpenalty=10000 
         \small \noindent \hangindent\parindent \hangafter=-1 
         \hbox to0pt{\hskip-\hangindent \warnsym\hfill}\ignorespaces}%
  {\par\endgroup\medbreak}

\title{Getting Started with \isajedit in \isayear}
\titlerunning{Getting Started with \isajedit}
\author{Christian Sternagel\thanks{%
This work is supported by the Austrian Science Fund (FWF) projects J3202 and
P27502.}}
\authorrunning{C.~Sternagel}
\institute{University of Innsbruck, Austria\\
  \email{christian.sternagel@uibk.ac.at}}

\begin{document}
\maketitle

\begin{abstract}
This is a 
beginner-oriented introduction to \isajedit,
the main user interface for the proof assistant \isa.
\end{abstract}

\section{Introduction}
Before we start, let me clarify my goals.
This document is intended as a short introduction to \isajedit (as shipped with
\isaversion) from a new user's perspective.

As an aside: this is an updated version of an earlier document (for
\texttt{Isabelle2012}), in which I also highlighted differences to the (now
obsolete) \pg interface.

In the remainder I will sometimes highlight common pitfalls for new users
(mostly issues that have been collected from the \isa mailing lists).
I do however \emph{not} give any technical or implementation details (for those,
I refer to \cite{Wenzel2010, Wenzel2011, Wenzel_Wolff2011, Wenzel2012,
Wenzel2012b} and the \isa documentation).
Moreover, my intended audience are newcomers to \isajedit (in other words: I
have nothing to say that people could not easily find out on their own, but hope
to save some people some time).


\begin{warn}
Points that may need special attention are marked like this.
\end{warn}

In the following, I use the following conventions: When indicating a menu like \menu{A\then
B\then C} I mean: ``Click on submenu \menu{C} in submenu \menu{B} of menu
\menu{A}.'' When indicating keyboard shortcuts like \scut{C-s} I mean: ``Keep
key \scut{C} pressed while pressing key \scut{s}.'' Here, \scut{C} refers
to the \emph{control} key (usually labeled with \textsf{Ctrl} on PC
keyboards; on Mac OS this is actually \textsc{command}).

The remainder is structured as follows:
\begin{itemize}
\item
First, in \secref{motiv}, I argue why
\isajedit is a great user interface for \isa.
\item
Then, in \secref{start}, I explain how to obtain and start \isajedit and give an
example walkthrough.
\item
The next section (\secref{use cases}) is
concerned with some typical use cases.
\item
Finally, I conclude in \secref{concl}.
\end{itemize}

\section{\label{sec:motiv}Why \isajedit is Awesome}

The most important point (in my opinion) is that \isajedit (or rather the
technology%
\footnote{\isajedit is a client to the PIDE framework \cite{Wenzel_Wolff2011};
there may be other clients (e.g., web-based or as part of an existing IDE, like
Eclipse).}
that makes it possible to use the editor \jedit as an interface to the
proof assistant \isa) provides an interaction model that is broadly known
and used nowadays (e.g., in word processors and integrated development
environments for programming languages). Let us call it the \emph{continuous model}.
The user sees and \emph{freely} edits a document and all the heavy
machinery of \isa asynchronously runs in the background and supplies
\emph{source text} with semantic information that is again presented in ways
people are used to nowadays (highlighting, hyperlinks, tooltips, wavy
underlines, etc.).

The traditional interaction model, in contrast (which is still employed by most
other proof assistants), is more like a command-line
interface, we sequentially enter commands (one at a time) and wait for the
corresponding output. Since this would be exceedingly tedious whenever we wanted
to change something we entered very early (essentially, we have to enter
everything again), this kind of interaction is usually combined with source
text. Such that instead of typing everything again, we just tell the system
which part of the source text has to be reloaded. However, in order for this to
work, we need some kind of \emph{locked region} (the part of the source text
that has already been processed) which we are not allowed to edit. Or
alternatively, reload everything after each change.


At this point you may wonder: ``What's the big deal about the continuous model?''
Since it may seem that the only difference to the traditional model (lets call
it the \emph{sequential model}) is that we do not have to manage a locked region
manually. Well, the main point is that the sequential model is inherently
sequential, whereas the continuous model offers plenty of space for
parallelization. Thus, \isajedit does not only provide a more modern (and for many
potential users more familiar) user interface but additionally facilitates
(again, the real reason is rather the underlying technology) to seamlessly make
use of multi-core architectures.


\section{\label{sec:start}Get Rolling}

Before we can use \isajedit, we have to download it. Since it is contained in the
official \isa release, we download that from:
\begin{quote}
\url{http://isabelle.in.tum.de} or
\url{http://www.cl.cam.ac.uk/research/hvg/Isabelle/} or
\url{http://mirror.cse.unsw.edu.au/pub/isabelle/}
\end{quote}
The correct version for your operating system and architecture should be
provided by the big green button and can be installed to an arbitrary directory
\isahome.

\begin{warn}
Do not try to combine arbitrary versions of PolyML, \isa, Scala, Java, and
\jedit if you want to avoid problems. \emph{Everything} you need is packaged in
the \isa bundle.
\end{warn}

To start \isajedit from a command line (e.g., under Linux) use (something similar to)
\begin{quote}\tt
\isahome/bin/isabelle jedit
\end{quote}

\begin{figure}
\begin{center}
\includegraphics{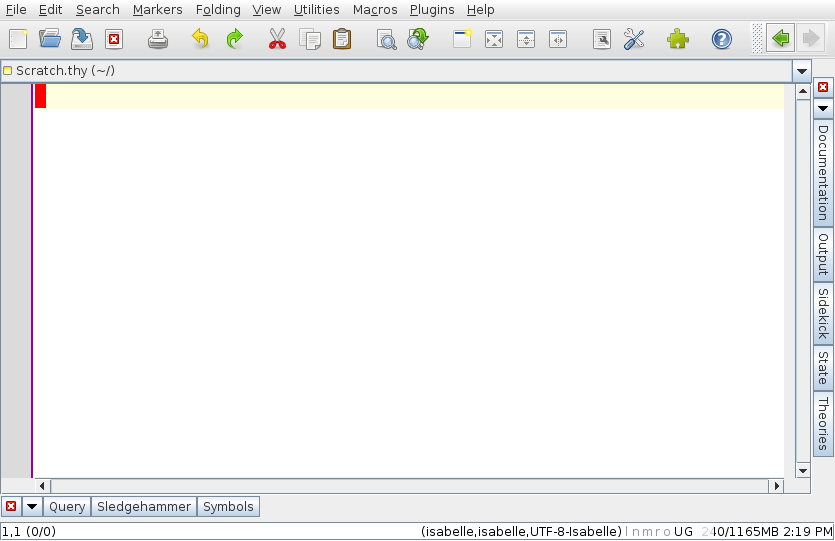}
\caption{\label{fig:first}\isajedit when first started.}
\end{center}
\end{figure}

In \figref{first} you see how \isajedit presents itself after being started the
first time (I just reduced the initial window size in favor of readability).
In the center
we have the \emph{main buffer}, this is where our source text is
shown and editing takes place.
On the right we have
(among
others) buttons \emph{Output} and \emph{Theories} which activate corresponding
panels when pressed.
When selecting \emph{Output} the panel shows the \emph{output buffer},
this is where messages from the \emph{current command} are to be found, where
the current command is determined by the cursor position in the main buffer
(that is, by default the output buffer shows the message(s) corresponding to the
command on which the cursor is positioned).

\begin{warn}
The \emph{Theories} panel can always be consulted to check whether \isajedit
agrees with you on which files are loaded (and how much of them), where problems
are indicated in \textcolor{red}{red}.
\end{warn}

\begin{warn}
\emph{Theories} also contains a drop-down list where it is possible to select a
certain logic image (the default is \emph{default (HOL)}). This does, however,
require a restart to take effect.
Alternatively you can start \isajedit via
\begin{quote}{\tt
isabelle jedit -l Image
}\end{quote}

\noindent
where \texttt{Image} is the name of the desired logic image.
\end{warn}

To start using \isajedit let us formalize a simple fact about lists. Before we can
do so, we need to start a \emph{theory} (which is \isa parlance for a source
file that bundles definitions and facts under a common name). We start by typing
``\type{\bl{theory} Test \gr{imports} Main \gr{begin}}'' into the main buffer,
which tells the system to start a theory with name \emph{Test} on top of the
theory named \emph{Main}, the default starting point for \isa theories
(that is, providing some kind of ``standard library'' of definitions and facts). At
this point you may notice an error message in the output buffer (which is
additionally stressed by showing a red sign in the left margin (the
\emph{gutter}) as well
as wavy underlining the word ``\type{Test}'' in the input.
\begin{quote}
\error{Bad theory name "Test" for file "Scratch.thy"}
\end{quote}
This happens because the default file name for theories is
\texttt{Scratch.thy} and can easily be resolved by saving (either \menu{File\then
Save} or \scut{C-s}) the current file under the name \texttt{Test.thy}.
\begin{warn}
A theory with name \emph{Name} has to reside in a file with name
\texttt{Name.thy}. The \isa convention is to capitalize theory names and
separate words by underscores, e.g., \texttt{Complete\char`_Partial\char`_Order}
rather than \texttt{completePartialOrder},
\texttt{complete\char`_partial\char`_order}, etc.
\end{warn}

Now, let us come back to the announced fact about lists: when we combine the
head and the tail of a non-empty list, we obtain the same list again. To make it
a bit more interesting (and informative), we define our own functions for
computing the head and the tail of a list as follows:
\begin{quote}
{\tt
\bl{fun} head \mbox{::} "\tvar{a} list => \tvar{a}" \gr{where}
  "\free{head} (\bound{x} \# \bound{xs}) = \bound{x}"}\\
{\tt
\bl{fun} tail \mbox{::} "\tvar{a} list => \tvar{a} list" \gr{where}
  "\free{tail} (\bound{x} \# \bound{xs}) = \bound{xs}"}\\
\end{quote}
There are already a few things to note here. When defining \texttt{head}, a
warning occurs (which is additionally stressed by showing a yellow sign in
the gutter as well as wavy underlining the word ``fun'' in the input)
\begin{quote}
\warning{Missing patterns in function definition:}\\
\warning{\free{head} [] = undefined}
\end{quote}
which tells us that our function is not defined on input \texttt{[]}. This is
not an error but just a warning, since we could intend our function to be
undefined in this case (as we indeed do). A similar warning is issued for
\texttt{tail}. Further note that the status of an identifier is indicated by its
color: free variables are rendered \free{blue}, bound variables \bound{green},
defined constants black, etc. Moreover, many entities allow us to jump to their
definition by holding \scut{C} pressed while left-clicking on them with the
mouse. For \type{\free{head}} and \type{\free{tail}} this means that we end up
in the line containing the corresponding \type{\bl{fun}} keyword (even if that
line is part of a different file). We can test this behavior by
\scut{C}-left-clicking \type{\#} in the definition of \type{\free{head}} which
brings us to the theory \emph{List} of \isa's standard library. More
concretely, to the line where the list datatype is defined (since \type{\#}
refers to one of the constructors of the list datatype). The fastest way of
going back is \scut{C-\textquotesingle} (control together with backtick, not to be
confused with a single quote) or \menu{View\then Go to Recent Buffer}.

Now we prove our little lemma, giving it the name
\texttt{head\char`_tail\char`_id} (note that at this point \type{head} and
\type{tail} are printed black, since they are defined constants now):
\begin{quote}{\tt
\bl{lemma} head\char`_tail\char`_id:~"\inot (\free{xs} = []) ==>
  head \free{xs} \# tail \free{xs} = \free{xs}" \\
 \mbox{}~~\bl{by} (cases \free{xs}) simp\char`_all
}\end{quote}

We finish our small theory by adding \type{\gr{end}} at the end of the file.


\section{\label{sec:use cases}How To \ldots}

In this section I collect typical use cases of things that you might want
(or have) to do inside \isajedit. It is organized by typical user
questions/statements.

\subsection{How can I use fancy mathematical symbols?}
This is really easy and convenient thanks to the code completion facility of the
SideKick plugin. By default, code completion is active (you can change the
corresponding settings at \menu{Plugins\then Plugin Options\ldots\then
SideKick}) and works as follows: you just enter what you mean in plain ASCII and
if available the system offers you a popup with alternative choices. If you want
to close the popup, just keep on typing (if you are at the end of a word just
enter a space) or press \scut{ESC}. If you want to choose an alternative,
navigate through the popup using the arrow keys (or the mouse) and press
\scut{TAB} when you are satisfied.
Some symbols will be completed automatically.  For example, after typing
\type{==>} (two equal signs followed by a greater than) you will end up with
$\Longrightarrow$.
Many symbols also have names that are close to the usual
\LaTeX{} names, e.g.,
type \type{\char`\\forall} (followed by \scut{TAB}) to obtain $\forall$,
\type{\char`\\exists} to
obtain $\exists$, etc.
Often, you do not even have to enter the full name,
a prefix is enough, e.g., after typing \type{\char`\\fo} you will obtain a choice
between the symbol $\forall$ and several other symbols.
\begin{warn}
An alternative way to enter ``Isabelle symbols'' is like \type{\char`\\<forall>}
(and this is how special symbols are actually stored in your theory files). 
\end{warn}

\section{\label{sec:concl}Conclusions}

I summarized some of the points that have already been discussed
on the \isa mailing lists (or elsewhere). Nevertheless, I hope that this
short and beginner-oriented introduction may help potential new users out there
to get up and running faster than they would have without this document.


\paragraph{Acknowledgments.}
I want to thank the \isa community (especially the developers) for always
being very helpful (and friendly) on the \isa mailing lists and Makarius
Wenzel for developing such an awesome tool as \isajedit (keep going!).

\bibliography{references}
\end{document}